\documentstyle[preprint,aps]{revtex}
\begin{document}

\draft
%{\tighten
\preprint{\vbox{\hbox{U. of Iowa preprint 96-13}}}

\title{NON-PERTURBATIVE ASPECTS OF SCALAR FIELD 
THEORY\footnote{Talk given by Y. Meurice at DPF 96, Minneapolis, 
August 1996}  }

\author{Y. Meurice, S. Niermann and G. Ordaz}

\address{Department of Physics and Astronomy\\
University of Iowa, Iowa City, Iowa 52246, USA}

\maketitle

\begin{abstract}
Using the hierarchical approximation, we 
discuss the cut-off dependence of the renormalized
quantities of a scalar field theory. The naturalness problem
and questions related to triviality bounds are briefly discussed.
We discuss unphysical features associated with the hierarchical
approximation such as the recently observed oscillatory 
corrections to the scaling laws.
We mention a two-parameter family of recursion formulas which
allows one to continuously extrapolate between Wilson's approximate recursion
formula and the recursion formula of Dyson's hierarchical model.
The parameters are the dimension $D$ and $2^{\zeta}$, the number of 
sites integrated in one RG transformation.
We show numerically that at fixed $D$,
the critical exponent $\gamma $ depends continuously on ${\zeta}$.
We suggest the requirement 
of $\zeta -$independence as a guide for constructing 
improved recursion formulas.
\end{abstract}

%\narrowtext
\newpage

One important unsolved problem of the electroweak interactions
is the mechanism responsible for the symmetry breaking.
With a single scalar doublet (Higgs), the standard model provides
an economical candidate solution. 
There are two issues related to scalar field theory which
are relevant to restrict and understand better the symmetry 
breaking mechanism. The first one is the so-called naturalness
(or fine-tuning, or hierarchy) problem \cite{bardeen}. The second
is that arbitrarily large values of $M_H / M_W$ are 
forbidden \cite{hazenfratz} when $\Lambda > M_H$, where $M_H$ is
the mass of the Higgs, $M_W$ the mass of the $W$ boson 
and $\Lambda $ a UV cutoff.

One possible method to address these questions non-perturbatively
consists in  using Wilson's approximate recursion formula \cite{wilson}.
In this approximation the renormalization group transformation
reduces to a simple integral equation closely related to
the one associated with Dyson's hierarchical model.
This method allows exact large volume calculations \cite{num}
which can be used to test numerical approximations.
On the other hand, the improvement \cite{marseille} 
of this type of approximation
is difficult.

The so-called naturalness problem of scalar field theory arises
when one tries to take the limit of large UV cutoff $\Lambda $
keeping the renormalized mass $m_R$ of the order of a fixed
cut-off of reference $\Lambda _R$. This can be done explicitly
using the approximate recursion formula with an initial potential
$Q_0(\phi)$. Applying $L$ times the renormalization group
transformation, one obtains a theory with $\Lambda _R= 2^{-L}\Lambda$
and a local potential $Q_L$.
Maintaining $m_R$ at a fixed not too large value 
in $\Lambda _R$ units requires a fine
tuning \cite{wilson} of the $r\phi ^2 $ term in $Q_0$. 
Namely, $r=r_c+A2^{-BL}$,
where $A$ depends on the choice of $m_R$ and $A$ takes the 
approximate values 1.6 in $D=3$ and 2.0 in $D=4$.
Note that for $D=3$, this result is not what would have guessed
using perturbation theory.
 
The fine-tuning implies practical restrictions.
To fix the ideas, if the 
calculation are done in double precision and if $\Lambda _R$ = 100 $GeV$
in $D$ = 4, one can only reach the value $\Lambda$ = $10^{10}$ $GeV$.
Is this a serious problem which could force a physical solution such
as supersymmetry or technicolor? There is no clear answer to this
question, however one should bear in mind that this might be 
an artifact of our calculational approach. An inspiring example might
the fine-tuning of the energy in a 
calculation \cite{anh} of the wave functions of the anharmonic oscillator 
far away from the origin. A question which can be answered is the 
question of the stability of the renormalized quantities
under the fine-tuning. When $\Lambda_R\ < \ m_R$, one can use $Q_L$, to 
calculate \cite{gust} the renormalized coupling constant at zero momentum 
non-perturbatively. The numerical value of the renormalized
quantities stabilizes well when $L$ becomes
large. Consequently, the fine-tuning is not incompatible with predictability.

Another important non-perturbative question is to find the forbidden
region in the ($M_H/M_W$, $\Lambda / M_H$) plane.
Numerical calculations \cite{hazenfratz} using the hierarchical
approximation or related approximations indicate an
upper limit $M_H\ < \ 10 M_W$ (triviality bound). In addition, if this bound
is reached, $\Lambda$ and $M_H$ must be of the same order. 
One might want to check the semi-classical formulas used in these calculations
without any reference to any perturbative expansion. In the symmetric
phase, we obtained an agreement within 1 percent. A similar check in the 
broken phase involves more subtle aspects and is presently in 
progress \cite{gust}.

The main source of errors in the triviality bounds mentioned above
is probably the use of the hierarchical approximation.
In particular, one needs to be aware of the unphysical features 
associated with it.
We calculated \cite{prl} 800 coefficients 
of the high-temperature expansion
of the magnetic susceptibility
of Dyson's hierarchical model with a 
Landau-Ginzburg
measure and a Ising measure in $D=3$.
Log-periodic corrections to the scaling laws appear in both cases.
The period of oscillation is
given in good approximation
by the logarithm of
the largest eigenvalue of the linearized RG transformation   
in agreement with a possibility suggested by K. Wilson \cite{wilson}.
We estimated $\gamma $ to be 1.300 (with a systematic error
of the order of 0.002)
in good agreement with the results
obtained with other methods such as the $\epsilon $-expansion \cite{epsi}.
The oscillations reflects a discrete scale invariance of the 
hierarchical model rather than a physical property of the model
approximated. The period would change if we could continuously
change the largest eigenvalue. This is the topic of the next
paragraph.

We found \cite{zeta} 
a two-parameter family of recursion formulas for scalar field
theory. The first parameter is the dimension $D$. The second parameter 
is $\zeta$, where $2^{\zeta}$ is the number of sites integrated 
by one renormalization group transformation.
Changing $\zeta$
allows one to continuously extrapolate between Wilson's approximate recursion
formula and the recursion formula of Dyson's hierarchical model.
We showed numerically that at fixed $D$,
the critical exponent $\gamma $ depends continuously on $\zeta$.
We would like to modify the recursion formula in order to 
get $\zeta $-independent physical quantities. 
Under some assumptions discussed in Ref. [10], one can construct
systematically ``counterterms'' that cancel the 
$\zeta$-dependence. This procedure might be somehow analogous to
the $\mu$-independence used in perturbative field theory \cite{gross}.

Finally, we would like to say a few words about the large $N$ limit
of scalar field theory. The recursion formula of Ref. [9]
can be easily extended to $N$ components. The leading term in the 
$1/N$ expansion of the initial probability for $O(N)$ sigma models
is Gaussian, however the high-temperature expansion has non-gaussian
features which are presently under study \cite{steve}.

\def\NCA{\em Nuovo Cimento}
\def\NIM{\em Nucl. Instrum. Methods}
\def\NIMA{{\em Nucl. Instrum. Methods} A}
\def\NPB{{\em Nucl. Phys.} B}
\def\PLB{{\em Phys. Lett.}  B}
\def\PRL{\em Phys. Rev. Lett.}
\def\PRD{{\em Phys. Rev.} D}
\def\PRB{{\em Phys. Rev.} B}
\def\ZPC{{\em Z. Phys.} C}

%} 

\vfil
\end{document}